\documentclass[
    ,final            % use final for the camera ready runs
  ]
  {aipproc}

\layoutstyle{6x9}

\begin{document}

\title{Monte Carlo approach to nuclei and nuclear matter}


\author{Stefano Fantoni}{address={S.I.S.S.A., International School of Advanced Studies},
altaddress={INFN, Sezione di Trieste and INFM, CNR--DEMOCRITOS National Supercomputing Center}}

\author{Stefano Gandolfi}{address={S.I.S.S.A., International School of Advanced Studies},
altaddress={INFN, Sezione di Trieste}}

\author{Alexey Yu. Illarionov}{address={S.I.S.S.A., International School of Advanced Studies},
altaddress={INFN, Sezione di Trieste}}

\author{Kevin E. Schmidt}{address={Department of Physics, Arizona State University}}

\author{Francesco Pederiva}{address={Dipartimento di Fisica, University of Trento},
altaddress={INFM, CNR--DEMOCRITOS National Supercomputing Center}}

\begin{abstract}
We report on the most recent applications of the Auxiliary Field Diffusion Monte
Carlo (AFDMC) method. 
The equation of state (EOS) for pure neutron matter in both normal and BCS phase 
and the superfluid gap in the low--density regime are computed,
using a realistic Hamiltonian containing the Argonne AV8' plus 
Urbana IX three--nucleon interaction.
Preliminary results for the EOS of isospin--asymmetric nuclear matter are also presented.
\end{abstract}

\maketitle

\section{Introduction}
The improved accuracy of experimental data on nuclei, together with a
rediscovered role of nuclear matter properties in the understanding of nuclear 
structure and several phenomena of 
astrophysical interest\cite{raffelt96}, has shown the need for
a more detailed investigation of the 
ground state of nuclear many--body systems.

Recent realistic nuclear Hamiltonians have been used to compute properties of light nuclei
in very good agreement with experiments\cite{pieper05}. However, the physical
properties of nuclear 
and neutron matter could be very different from that of nuclei; in fact the nucleon density 
in the core of heavy nuclei reaches the maximum value of $\rho_0$=0.16 fm$^{-1}$,
while the relevant range of density of matter inside neutron stars is up to 9 times $\rho_0$\cite{piekarewicz04}.
Therefore we are now facing, on one side, the problem of finding a fundamental scheme for 
the description of nuclear forces, valid from the deuteron up to dense nuclear matter, 
which is still an open fundamental problem,
and, on the other, that of solving a many--body problem which is made extremely complex 
by the strong spin--isospin dependence of the forces. 
In this paper we will not address the problem of determining
the nuclear force. We will consider a nuclear Hamiltonian
which provides good fits to the N--N data up to meson production and reproduces
fairly well the ground state and the low energy spectra of light nuclei. This is 
made of two- and three-body spin--isospin dependent potential. We
present results for the ground state of large nuclear systems
with this Hamiltonian.

It is well known that Quantum Monte Carlo (QMC) methods can
provide estimates of physical observables 
at the best known accuracy\cite{schmidt84}, and they are therefore useful to 
gauge the validity of proposed interaction models without having the bias of using more approximate methods. 

A new generation of powerful QMC techniques have been recently devised to simulate
large nucleonic 
systems with up to hundred of nucleons: the Auxiliary Field Diffusion 
Monte Carlo (AFDMC)\cite{schmidt99}. They have been used to compute the EOS of nuclear 
matter\cite{gandolfi07}, showing important limitations of other many--body methods and of the 
modern nuclear interactions based on two-- plus three--body potentials in the high density regime.
The accuracy of AFDMC was demonstrated by comparing the ground state of light nuclei with 
results provided by Green's Function Monte Carlo (GFMC)\cite{pudliner97} that is
known to give accurate results for the properties of light nuclei up to 
A=12\cite{pieper05}.
By sampling the spin--isospin states of the nucleons AFDMC can be applied
to large systems; it was used to simulate the ground state of medium sized
nuclei 
up to A=40\cite{gandolfi07b}, nuclear matter with up to A=108\cite{gandolfi07}, 
the properties of neutron--rich nuclei\cite{gandolfi06,gandolfi08},
neutron drops\cite{gandolfi07c}, and neutron matter with up to A=114\cite{gandolfi07c,sarsa03,gandolfi08c}.

In this paper we discuss the latest results of the computation of the equation of state (EOS) of
neutron matter in both the normal and superfluid phases, the computation of the superfluid 
gap of neutron matter in the low--density regime, and some preliminary results about the EOS of
isospin-asymmetric nuclear matter.

\section{Hamiltonian}
The ground state of nuclear systems can be realistically studied by starting from the non--relativistic
nuclear Hamiltonian of the form
\begin{equation}
H=-\frac{\hbar^2}{2m}\sum_{i=1}^N\nabla_i^2+\sum_{i<j}v_{ij}+\sum_{i<j<k}V_{ijk} \,,
\end{equation}
where $m$ is the averaged mass of proton and neutron, and $v_{ij}$ and $V_{ijk}$ are 
two-- and three--body potentials; it seems that the effect of forces due to n--body 
terms with $n>3$ in the low energy properties of light nuclei is negligible.
Such a form for the Hamiltonian has been shown to describe properties of light nuclei in excellent
agreement with experimental data (see Ref. \cite{pieper05} and references therein).
All the degrees of freedom responsible for the interaction between nucleons (such the $\pi$,
$\rho$, $\Delta$, etc.) are integrated out and included in $v_{ij}$ and $V_{ijk}$.

At present, several realistic nucleon--nucleon interactions (NN) fit scattering data with very
high precision. We consider the NN potentials belonging to the Argonne family.
Such interactions are written as a sum of operators:
\begin{eqnarray}
v_{ij} = \sum_{p=1}^{M} v_p(r_{ij}) O^{(p)}(i,j)\,,
\end{eqnarray}
where $O^{(p)}(i,j)$ are spin--isospin dependent operators.
The number of operators $M$ characterizes the interaction; the most accurate of them is 
the Argonne AV18 with M=18\cite{wiringa95}. Here we consider some simpler forms
derived from AV18, namely the AV8' and the AV6'\cite{wiringa02} with a smaller 
number of operators. For many systems, the difference between these simpler
forms and the full AV18 potential can be computed perturbatively.
Most of the contribution of the NN is due to the one pion exchange between nucleons, but
the effect of other mesons exchange as well as some phenomenological terms are included.

The eight $O^{(p)}(i,j)$ terms in AV8' are given by the four central components $1$,
$\vec\tau_i\cdot\vec\tau_j$, $\vec  \sigma_i \cdot \vec \sigma_j$, 
$(\vec \sigma_i \cdot \vec \sigma_j)(\vec \tau_i\cdot \vec \tau_j)$,
the tensor $S_{ij}$, the tensor--$\tau$ component $S_{ij} \vec \tau_i\cdot \vec \tau_j$,
where $S_{ij} = 3 (\vec \sigma_i \cdot \hat r_{ij}) (\vec \sigma_j \cdot
\hat r_{ij}) -\vec \sigma_i \cdot \vec \sigma_j$, the spin--orbit $\vec L_{ij}\cdot\vec S_{ij}$ 
and the spin--orbit--$\tau$ $(\vec L_{ij}\cdot\vec S_{ij})(\vec\tau_i\cdot \vec \tau_j)$,
where $\vec L_{ij}$ and $\vec S_{ij}$ are the total angular momentum and the total spin of the pair $ij$.

The AV6' has the same structure of AV8', but the spin--orbit operators are dropped.
In general, all the AVx' interactions are obtained starting from the AV18, written by dropping
less important operators, and \emph{refitted} in order to keep the most important features of NN 
in the scattering data\cite{wiringa02}.

The three--body interactions (TNI) is essential to overcome the underbinding of 
nuclei with more than two nucleons. The NN is fitted to scattering data and correctly 
gives the deuteron binding energy, but starting with $^3H$ the NN is not sufficient
to describe the ground state of light nuclei.
The Urbana-IX (UIX) potential corrects this limitation of NN, and was fitted to light nuclei and to 
correctly reproduce the expected saturation energy of nuclear matter\cite{pudliner95}.
It essentially contains the Fujita--Miyazawa term\cite{fujita57} that describes the 
exchange of two pions between three nucleons, with the creation of an intermediate 
excited $\Delta$ state.
Again, a phenomenological part is required to sum all the other neglected terms.
The generic form of UIX is the following:
\begin{equation}
V_{ijk}=V_{2\pi}+V_R \,.
\end{equation}
The Fujita-Miyazawa term\cite{fujita57} is spin--isospin dependent:
\begin{equation}
V_{2\pi}=A_{2\pi}\sum_{cyc}\Big[\{X_{ij},X_{jk}\}\{\tau_i\cdot\tau_j,\tau_j\cdot\tau_k\}+
\frac{1}{4}[X_{ij},X_{jk}][\tau_i\cdot\tau_j,\tau_j\cdot\tau_k]\Big] \,,
\end{equation}
where the $X_{ij}$ operators describe the one pion exchange, and their structure is 
the same of that of AV6'.
The phenomenological part is
\begin{equation}
V_{ijk}^R=U_0 \sum_{cyc}T^2(m_\pi r_{ij})T^2(m_\pi r_{jk}) \,.
\end{equation}

The factors $A_{2\pi}$ and $U_0$ are kept as fitting parameters. The binding energy of symmetrical 
nuclear matter is not well reproduced by such force.
Other forms of TNI, called Illinois forces\cite{pieper01}, which includes three--nucleon Feynman diagrams
with two Deltas intermediate states, are available.
However, they provide unrealistic overbinding of neutron systems when density 
increases\cite{gandolfi07c,sarsa03} 
and they do not seem to describe realistically high density (already at $\rho\ge\rho_0$) nucleonic systems.
 
\section{The AFDMC Method}
Ground state AFDMC simulations rely, as do other traditional QMC methods, on previous variational 
calculations, often performed within FHNC theory\cite{pandharipande79}, to compute a trial wave 
function $\Psi_T$, which is used to guide the sampling of the random walk. A typical form 
for $\Psi_T$ is given by a correlation operator $\hat F$ operating on a mean field wave function $\Phi(R)$,
\begin{eqnarray}
\langle R, S |\Psi_T\rangle = \hat{F} \Phi(R)\,.
\end{eqnarray}

Mean field wave functions $\Phi(R)$ that have been used are: (i) a Slater determinant $\Phi_{FG}$ 
of plane wave orbitals for nuclear and neutron matter in the normal phase, (ii) a linear 
combination $\Phi_{sp}$ of a small number of antisymmetric products of single particle 
orbitals $\phi_j(\vec r_i,s_i)$  for nuclei and neutron drops, and (iii) a pfaffian 
$\Phi_{pf}$, namely an antisymmetric product of independent pairs for neutron matter in superfluid phase.

A realistic correlation operator is the one provided by FHNC/SOC theory, namely $\cal S$$ \prod_{j>i} \sum_{p=1}^M f^{(p)}(r_{ij}) O^{(p)}(i,j) $, where $\cal {S} $ is the symmetrizer and the operators $O^{(p)}(i,j) $ are the same as those appearing in the two--body potential.

Unfortunately, the evaluation of
this wave function requires exponentially increasing
computational time with the number of particles.
This procedure is followed in variational and Green's function Monte Carlo calculations,
where the full sum over spin and isospin degrees of freedom is carried out.
Since for large numbers of particles it is not computationally
feasible to evaluate these trial functions,
the much simpler correlation operator $\prod_{j > i} f^c(r_{ij})$, which contains the central Jastrow
correlation only, is used instead. The evaluation of the corresponding trial function requires order $A^3$
operations to evaluate the Slater determinants and $A^2$ operations for
the central Jastrow. Since many important correlations are neglected in
these simplified functions, we use the Hamiltonian itself to define the
spin sampling.

The AFDMC method works much like Diffusion Monte Carlo\cite{schmidt84,schmidt99,
fantoni01,pederiva04,gandolfi06}. The wave
function is defined by a set of what we call walkers. Each walker is a set
of the $3A$ coordinates of the particles plus a number
$A$  of four component spinors each representing a spin--isospin state.
The imaginary time propagator for the
kinetic energy and the spin--independent part of the potential is identical
to that used in standard diffusion Monte Carlo. The new positions are sampled from
a drifted Gaussian with a weight factor for branching given by the local
energy of these components. Since they do not change the spin state,
the spinors will be unchanged by these parts of the propagator.

To sample the spinors we first use a Hubbard Stratonovich transformation to
write the propagator as an integral over auxiliary fields of a separated
product of single particle spin--isospin operators. We then sample the
auxiliary field value, and the resulting sample independently
changes each spinor for each particle in the sample, giving a new sampled
walker.

More details about the AFDMC method can be found in Ref. \cite{gandolfi07c}.

\section{Nucleonic Matter}

The properties of nuclear matter, like the Equation of State (EOS), are of fundamental 
importance in nuclear physics, mainly because nuclei behave very much like liquid 
drops.
Indeed, each of these can be associated with a mass formula, which fits the corresponding 
data of stable nuclei from  $A\sim 20$ on.
Any such mass formula has a volume and a symmetry term provided by symmetrical nuclear matter 
and nuclear matter with $N>Z$ respectively.
Moreover, accurate model independent calculations of the above observables are much needed 
in the physics of heavy ion reactions, as well as in that of lepton and neutrino scattering 
off nuclei at intermediate energies. Medium effects have to be taken into account for the 
data analysis of such reactions at the present level of accuracy.

In addition, the theoretical knowledge of the properties of asymmetric nuclear matter at 
low temperature is needed to predict the structure, the dynamics and
the evolution of stars, in particular during their last stages, when they become
ultra--dense neutron stars.

We present in the following the results obtained with AFDMC for the EOS of 
pure neutron matter in normal phase, as well as the gap of the BCS phase of 
neutron matter\cite{gandolfi07c,gandolfi08b,gandolfi08c}. Previously results 
of the EOS of nuclear matter and nuclei can be found in Refs. \cite{gandolfi07,gandolfi07b,gandolfi07c}.

%\subsection{Neutron matter}
Neutron matter is simulated by considering $N$ neutrons in a periodic box, and particular care 
is taken to evaluate the effects due to the finite size of the box. More details can be found in Refs.
 \cite{sarsa03,gandolfi07c}

\begin{figure}
\vspace{1.8cm}
\includegraphics[width=7cm]{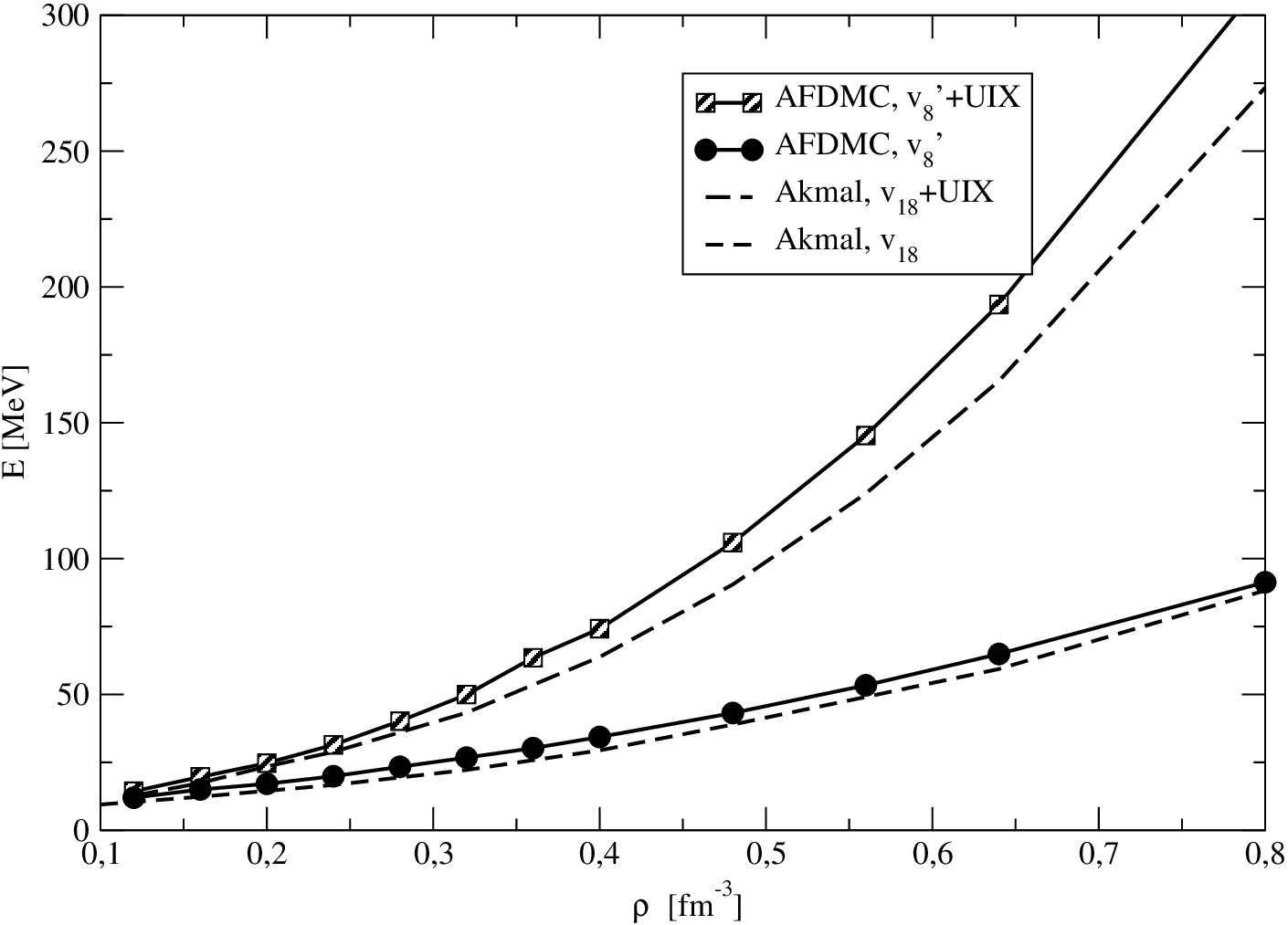}
\vspace{0.5cm}
\caption{The AFDMC equation of state evaluated by simulating 66 neutrons in a periodic box.
See the text for details.}
\label{fig:PNMeos}
\end{figure}

In Fig. \ref{fig:PNMeos} we plot
the AFDMC equation of state, obtained with the energy of 66 neutrons, and the variational 
calculation of Akmal et al. of Ref. \cite{akmal98}, where the AV18 NN interaction combined with the 
Urbana UIX TNI was considered. 
As it can be seen both the AV8' and the AV18 essentially give an EOS with the same behavior, but the addition 
of the TNI adds some differences, in particular at higher densities. 

The AV8' interaction should be more attractive than AV18 as shown in light nuclei and in neutron drop 
calculation\cite{pudliner97}. This result is not confirmed by our calculations as it is clearly visible in 
Fig. \ref{fig:PNMeos}. The AFDMC has proved 
to be in very good agreement with the GFMC results for light nuclei\cite{gandolfi07b}, and we believe that 
the same accuracy is reached in the neutron matter calculation, as shown in the comparison with 
the GFMC results of 14 neutrons\cite{gandolfi07c,gandolfi08c}. 
On the other hand the AFDMC calculation of the nuclear matter have shown that FHNC/SOC
does not seem to provide safe energy upperbounds. This is because of the lack of cluster diagrams with
 commutator terms beyond the SOC approximation and that of elementary diagrams,
not included in the FHNC summation\cite{gandolfi07}.

The addition of the UIX three--body interaction
to the Hamiltonian increases the differences between the AFDMC results and 
that of Akmal et al. The difference cannot be due to finite--size effects in our calculation for the 
following reason: the total contribution of UIX should be positive in neutron matter, so that the inclusion 
of box corrections as done for the two--body part of the Hamiltonian would eventually increase the 
total energy. The periodic box--FHNC estimation of these effect essentially confirms this observation\cite{sarsa03}.

It is worth observing how important the three--nucleon interaction already is
at medium--high densities. 
Its contribution at $2\rho_0$ is $\sim$ 25MeV and increases very rapidly with density.
The four Illinois potentials\cite{pieper01}, built to include two $\Delta$ intermediate states in the 
three nucleon processes, lead to very different results compared to the Urbana IX\cite{sarsa03,gandolfi07c} EOS at 
medium--high densities, in spite of the fact that all of them provide a satisfactory fit to the ground 
state and the low energy spectrum of nuclei with $A\leq 8$.   
This, once more, points outs the importance 
of understanding the role of $n$--body forces with $n > 3$ in nuclear astrophysics.

\begin{figure}%[ht]
\vspace{1.8cm}
\includegraphics[width=7cm]{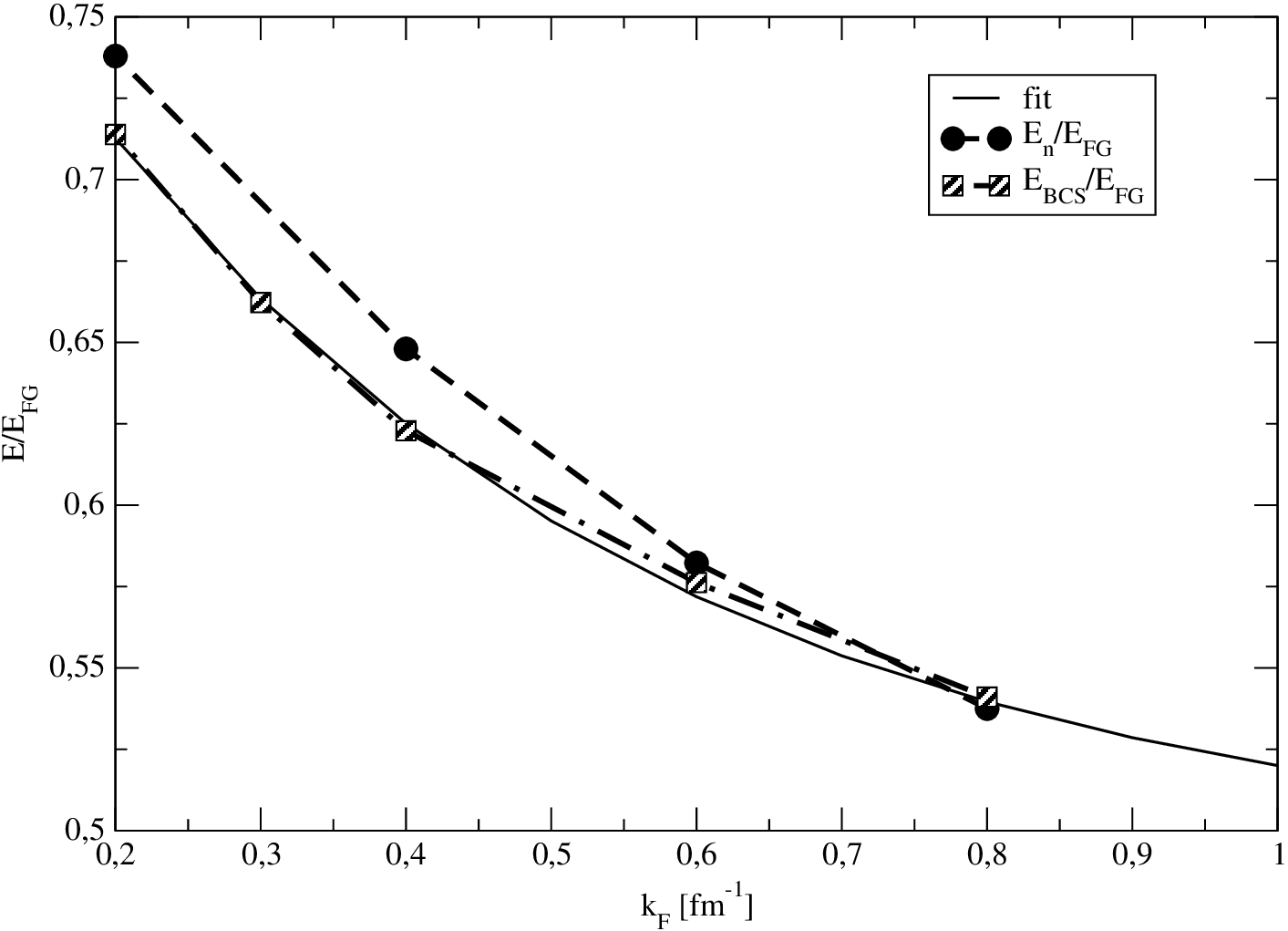}
\vspace{0.5cm}
\caption{The EOS of neutron matter in the low--density regime. The two calculations were performed 
using different trial wave functions modeling a normal and a BCS state. }
\label{fig:bcsEOS}
\end{figure}

We explored the superfluid phase of low--density neutron matter. Because the AFDMC 
projects out the lowest energy state with the same symmetry and phase as the trial wave function, we tried to repeat some
calculation using a BCS trial wave function of the form of Ref. \cite{fabrocini05}. The BCS state provides an energy 
that is lower than the normal state, however the difference never exceeds 3\% of the total energy.
We report the energy per neutron in the low--density regime evaluated using a normal and a BCS trial wave function
in Fig. \ref{fig:bcsEOS}.
 
In the low--density regime the neutron matter is in a $^1S_0$ superfluid phase. In this regime, the neutron--neutron 
interaction is dominated by this channel, whose scattering length is very large and negative, about a=--18.5 fm.
Several attempts to compute the pairing gap using a bare effective interaction
like those used for cold atom
problems have been
performed in the last few years\cite{gandolfi08b}. However, our results show that
an accurate calculation of the superfluid gap in this regime must include the full Hamiltonian instead 
of one describing the $^1S_0$ physics only. 

\begin{figure}%[ht]
\vspace{1.8cm}
\includegraphics[width=7cm]{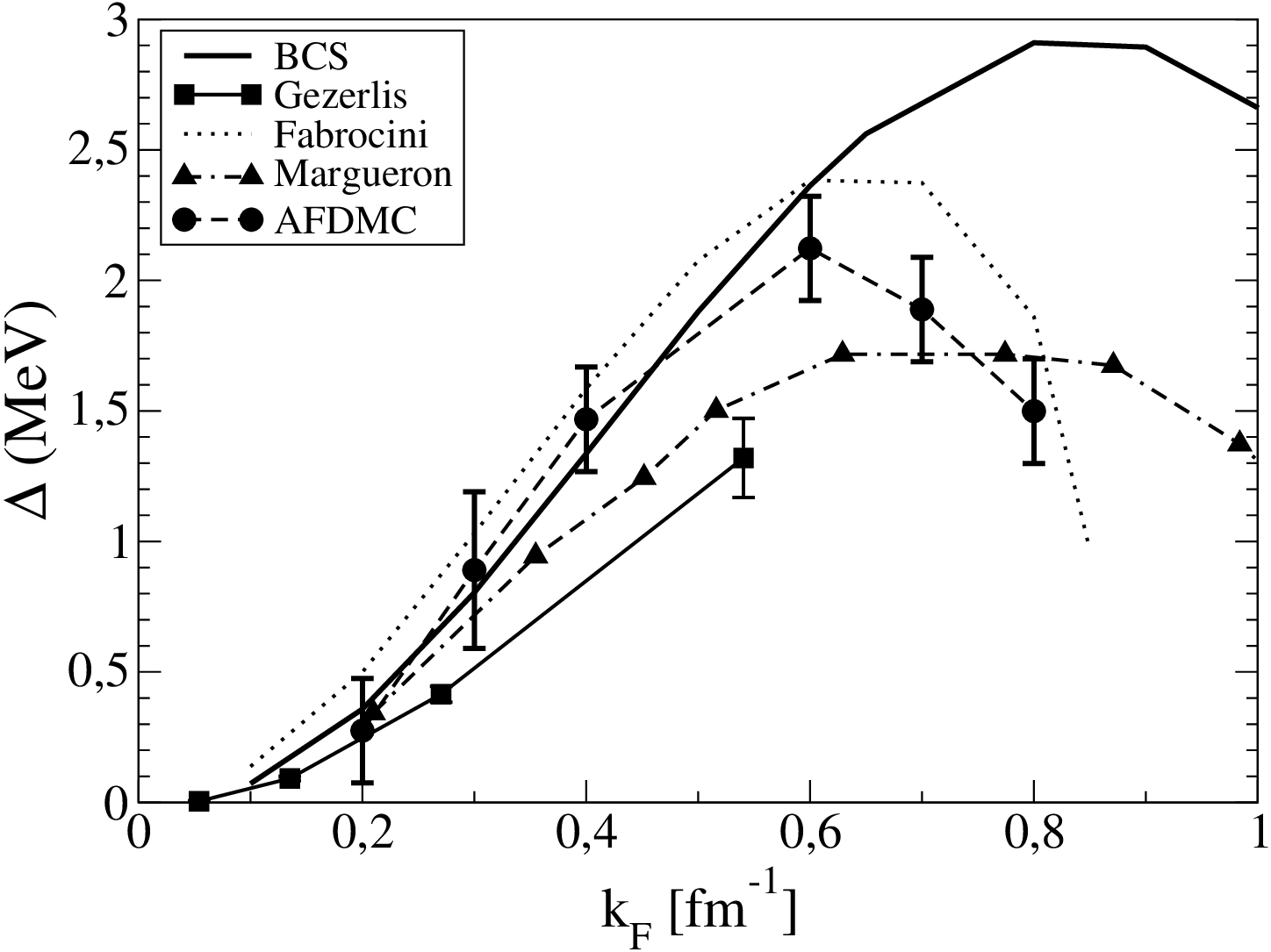}
\vspace{0.5cm}
\caption{The superfluid gap computed using AFDMC and compared with other techniques is displayed.}
\label{fig:gap}
\end{figure}

We computed the superfluid gap by means of AFDMC that allows for quantum simulations of the superfluid phase 
of neutron matter, by solving the ground state of the full Hamiltonian with AV8'+UIX without the use of 
some simplified interaction.
In Fig. \ref{fig:gap} the AFDMC gap is compared with standard BCS theory and some other often referred 
calculations based upon to correlated theories at two--body approximation (see also Ref. \cite{gandolfi08c}).

\begin{figure}%[ht]
\vspace{1.8cm}
\includegraphics[width=7cm]{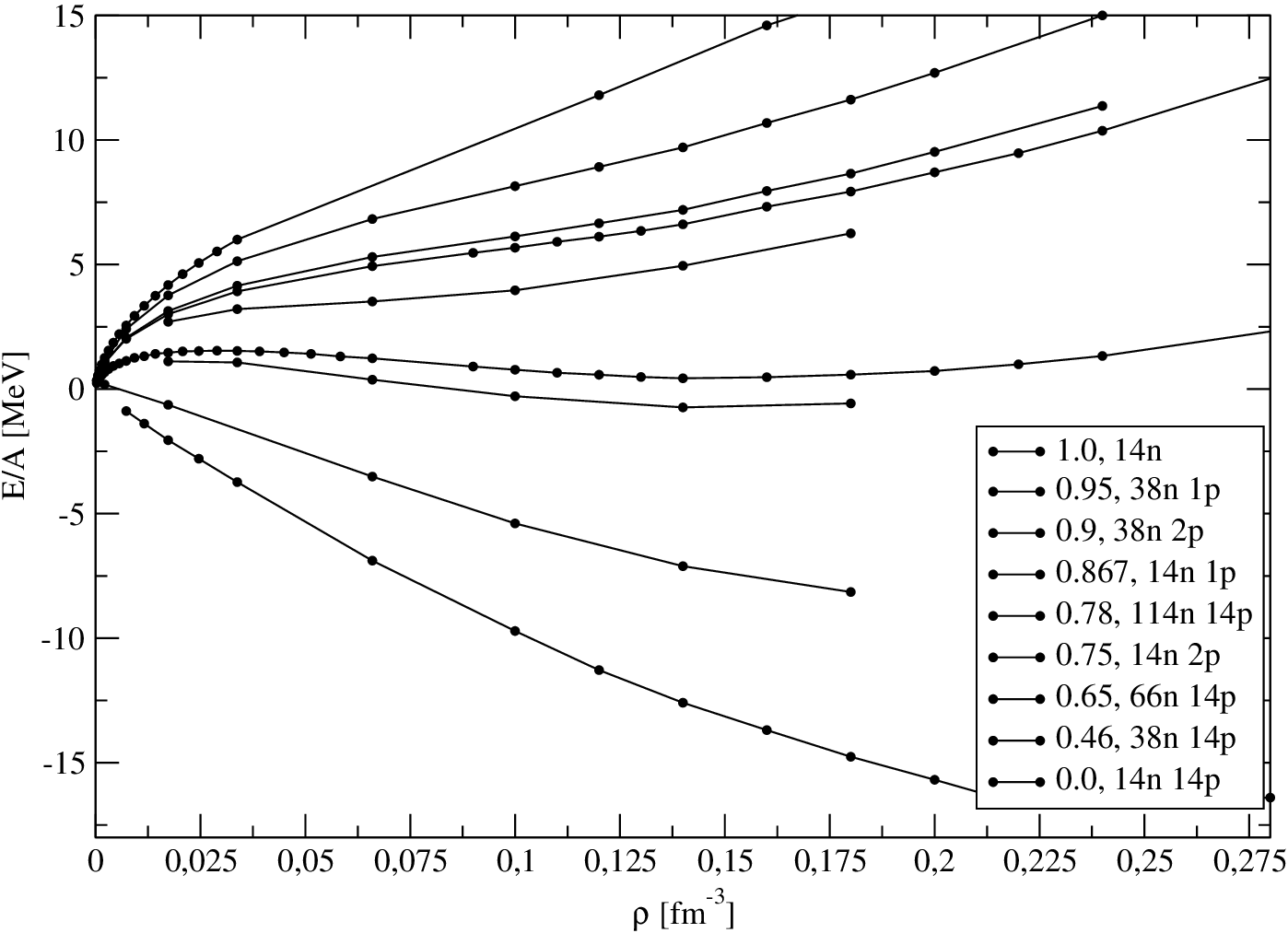}
\vspace{0.5cm}
\caption{The EOS of nuclear matter as a function of the density and of the isospin--asymmetry. See the text for details.}
\label{fig:asym}
\end{figure}

%\subsection{Nuclear Matter}
The computation of the isospin--asymmetric nuclear matter is possible with a trivial modification of the trial 
wave function used to project out the ground state of the system. The asymmetry is defined by
\begin{equation}
\alpha=\frac{N-Z}{N+Z} \,,
\end{equation}
where $N$ is the number of neutrons and $Z$ of protons.
We report in Fig. \ref{fig:asym} preliminary results of the EOS of nuclear matter for different values 
of $\alpha$, using a simplified Hamiltonian containing the AV6' NN interaction. The various curves of the 
figure from the top to the bottom are in the same order of the legend, where it is also indicated  
the number of neutrons and protons for each simulation.

As displayed in the legend of Fig. \ref{fig:asym}, some calculations were performed using a very small 
number of protons. These results could suffer important finite size effects. The study of finite 
size effect particularly due to the kinetic energy is in progress.
It is possible to reduce the dependence of the kinetic energy on the number of particles by using
different kinds of periodic boundary conditions: e.g. the so called twist average boundary conditions
(TABC)\cite{lin01}. The computation of several EOS by using TABC to see the effect to the kinetic energy is in progress.

\section{Conclusions and perspectives}
We have briefly shown some recent results obtained using AFDMC theory. 
This recently developed Quantum Monte Carlo provides results for the
binding energy of light nuclei\cite{gandolfi07b} and neutron drops\cite{gandolfi07c} which are is very good agreement 
with accurate GFMC calculations. However, unlike those methods
AFDMC is well suited to deal with nucleonic systems with up to hundred of nucleons in the Hamiltonian.

We have presented AFDMC results for the EOS of neutron matter in the normal phase\cite{gandolfi08c}. 
In the low--density regime where neutrons form a superfluid phase, we modified 
the trial wave function in order to include important BCS correlations to compute the superfluid gap of 
neutron matter\cite{gandolfi08b}.
Preliminary results for the EOS of asymmetric nuclear matter have also been presented and discussed. 

It should be stressed that, besides having shown that AFDMC theory opens up the possibility
of studying the properties of large nucleonic systems, with an accuracy which goes much beyond that of 
other commonly used many--body theories, the results we have already obtained indicate serious inadequacies
of commonly used nuclear interactions in the high density regime, of interest in astronuclear physics. 

The problem of determining the nuclear Hamiltonian, and in particular the effect of $n$--body forces with $n > 3$
is becoming of primarily importance. We are working on developing
a new form for the interaction that perturbatively contains the excitation of nucleons.
The corresponding potential naturally generates many-body forces, that should be
small in nuclei, but of primary importance in nuclear and neutron matter.

\begin{theacknowledgments}
We thank J. Carlson for very useful and stimulating discussions.
Calculations were partially performed on the BEN cluster at ECT* in Trento, under a grant
for supercomputing projects, and partially on the HPC facility of Democritos/SISSA.
\end{theacknowledgments}



\bibliographystyle{aipproc}   % if natbib is available

\bibliography{biblio}

\IfFileExists{\jobname.bbl}{}
 {\typeout{}
  \typeout{******************************************}
  \typeout{** Please run "bibtex \jobname" to optain}
  \typeout{** the bibliography and then re-run LaTeX}
  \typeout{** twice to fix the references!}
  \typeout{******************************************}
  \typeout{}
 }

\end{document}